\documentclass[twocolumn,showpacs,aps,amsmath,amssymb,prl,superscriptaddress]{revtex4-2}
\usepackage[latin9]{inputenc}
\usepackage{textcomp}
\usepackage{amsmath}
\usepackage{graphicx}
\PassOptionsToPackage{version=3}{mhchem}
\usepackage{mhchem}

\makeatletter

\providecommand{\tabularnewline}{\\}

\usepackage{subfigure}\usepackage{epstopdf}\usepackage{dcolumn}
\usepackage{bm}
\usepackage{multirow}
\usepackage{longtable}\usepackage{rotating}\usepackage{threeparttable}
\usepackage{float}
\def\br{{\bf r}}

\makeatother

\begin{document}
\title{Approximate Functionals in Hypercomplex Kohn-Sham Theory}
\author{Neil Qiang Su}
\email{nqsu@nankai.edu.cn}
\affiliation{Department of Chemistry, Key Laboratory of Advanced Energy Materials Chemistry (Ministry of Education) and Renewable Energy Conversion and Storage Center (RECAST), Nankai University, Tianjin 300071, China}

\begin{abstract}
The recently developed hypercomplex Kohn-Sham (HCKS) theory shows great potential to overcome the static/strong correlation issue in density functional theory (DFT), which highlights the necessity of further exploration of the HCKS theory toward better handling many-electron problem. This work mainly focuses on approximate functionals in HCKS, seeking to gain more insights into functional development from the comparison between Kohn-Sham (KS) DFT and HCKS. Unlike KS-DFT, HCKS can handle different correlation effects by resorting to a set of auxiliary orbitals with dynamically varying fractional occupations. These orbitals of hierarchical correlation (HCOs) thus contain distinct electronic information for better considering the exchange-correlation effect in HCKS. The test on the triplet-singlet gaps shows that HCKS has much better performance as compared to KS-DFT in use of the same functionals, and the systematic errors of semi-local functionals can be effectively reduced by including appropriate amount of the HCO-dependent Hartree-Fock (HF) exchange. In contrast, KS-DFT shows large systematic errors, which are hardly reduced by the functionals tested in this work. Therefore, HCKS creates new channels to address to the strong correlation issue, and further development of functionals that depend on HCOs and their occupations is necessary for the treatment of strongly correlated systems. 
\end{abstract}
\maketitle

%
%
%
%
%

\section{Introduction}

Density functional theory (DFT) \cite{HK1964,Levy1979pnas,KS1965,PY1989} is now the leading electronic structure method for both molecular and extended systems. Its success stems from the superiority of taking the electron density as the basic variable: the Hohenberg-Kohn theorem guarantees the theoretical accuracy, while the avoidance of solving the Schr{\"o}dinger equation renders the computational efficiency. Nonetheless, a DFT calculation may still deliver a qualitatively incorrect electron density or energy \cite{Cohen2008science,Chai2012jcp,Su2018pnas,Lee2019prl}. Thereby, a feasible method for obtaining electron density and a reliable functional for calculating electron energy are essential for the wider application of DFT.

Kohn-Sham (KS) DFT \cite{KS1965,PY1989,Dreizler2012,Mardirossian2017} is the commonly used realization of DFT. It introduces an easy-to-solve noninteracting reference system (i.e. the KS system), whose ground-state wave function is just a determinant constructed by occupied KS orbitals (i.e. the KS determinant). The special feature of the KS system is that, as compared to the physical interacting system, it yields the same electron density and exactly considers most of the kinetic energy. Therefore, KS-DFT provides an rigorous theoretical framework to calculate the electron density and energy, with only a small part of the total energy, that is, the exchange-correlation (XC) energy, being obtained by approximation.  While existing approximate functionals have made the great popularity of KS-DFT, the notorious failures in practical applications \cite{Ruzsinszky2006jcp,Ruzsinszky2007jcp,Vydrov2007jcp,Paula2008prl,Paula2009prl,Cohen2008science,Zheng2011},  especially the strong correlation issue \cite{Burke2012jcp,Cohen2012cr,Becke2014jcp,Su2017arpc,Perdewe2021pnas}, have severely reduced the reliability of KS-DFT calculations. 

The recently developed hypercomplex KS (HCKS) theory \cite{su2021unity} provides a promising alternative to realize DFT calculation. The generalization of the KS orbitals and the KS determinant to hypercomplex number systems leads to a set of auxiliary orbitals with fractional occupations, and formulates the kinetic energy of the reference system and the electron density with the auxiliary orbitals and their occupations. Exact in principle, HCKS relies on the approximation to the unknown XC energy in practice. Preceding tests in use of semi-local density functionals already demonstrated that HCKS is able to capture the multi-reference nature of strong correlation with dynamically varying fractional occupations \cite{su2021unity}. In addition, the auxiliary orbitals and their occupations provide extra flexibility, in addition to the electron density, in constructing approximate functionals toward better consideration of the XC effect in HCKS. Therefore, HCKS creates new channels for the development of DFT, which has the potential to overcome the limitation of KS-DFT in strongly correlated systems. 

Based on the theoretical framework of HCKS established in Ref \cite{su2021unity}, this work further discusses approximate functionals in HCKS, seeking to gain in-depth insights into the development of approximate functionals through the comparison between KS-DFT and HCKS. The advances in this work should be helpful for both the development and applications of DFT.

\section{Theory}

The Hohenberg-Kohn theorem states that the total electronic energy is uniquely determined by the electronic density \cite{HK1964}, such that 
\begin{equation}
\label{eq:DFT}
E_{\rm{tot}}[\rho_\sigma]=T[\rho_\sigma]+E_{\rm{ext}}[\rho_\sigma]+E_{\rm{ee}}[\rho_\sigma],
\end{equation}
which includes the kinetic energy $T[\rho_\sigma]$, the external energy $E_{\rm{ext}}[\rho_\sigma]$, and the electron-electron interaction energy $E_{\rm{ee}}[\rho_\sigma]$. Nevertheless, the exact density functional of the total energy is unknown, except $E_{\rm{ext}}[\rho_\sigma]$ and the Coulomb energy $E_{\rm{H}}[\rho_\sigma]$ (the classical contribution to $E_{\rm{ee}}[\rho_\sigma]$) that can be well-defined by  
\begin{equation}
\label{eq:ExtKS}
E_{\rm{ext}}[\rho_\sigma]=\sum_{\sigma}^{\alpha,\beta}\int v_{\rm{ext}}(\br)\rho_\sigma(\br)d\br,
\end{equation}
and
\begin{equation}
\label{eq:HKS}
E_{\rm{H}}[\rho_\sigma]=\frac{1}{2}\sum_{\sigma\sigma'}^{\alpha,\beta}\int\frac{\rho_\sigma(\br_1)\rho_{\sigma'}(\br_2)}{r_{12}}d\br_1d\br_2.
\end{equation}
It is thus inevitable to introduce an approximate treatment of $T[\rho_\sigma]$ and the non-classical part of $E_{\rm{ee}}[\rho_\sigma]$ (i.e. $E_{\rm{ee}}[\rho_\sigma]-E_{\rm{H}}[\rho_\sigma]$). However, the unknown energy are of large magnitude, especially the whole kinetic energy $T[\rho_\sigma]$, whether it is handled properly determines the practicality of DFT. In order to have a good treatment of $T[\rho_\sigma]$, the resort to orbitals (i.e. one-electron wave functions) can yet be regarded as a favorable choice, which can accurately consider most of $T[\rho_\sigma]$ with an acceptable computational cost, such as KS-DFT and HCKS.

\subsection{Theory and functionals of KS-DFT}

KS-DFT introduces a noninteracting reference system, whose ground-state wave function is just a determinant \cite{KS1965}, that is, the KS determinant constructed by the occupied KS orbitals, 
\begin{eqnarray}
\label{eq:wf_ks}
\Phi\!=\!\left|\!\begin{array}{cccccc} 
    \varphi_1^\alpha\!(\!\br_{\!1}\!)\alpha\! & \varphi_2^\alpha\!(\!\br_{\!1}\!)\alpha\! & \cdots & \varphi_1^\beta\!(\!\br_{\!1}\!)\beta\! & \varphi_2^\beta\!(\!\br_{\!1}\!)\beta\!   & \cdots \\ 
    \varphi_1^\alpha\!(\!\br_{\!2}\!)\alpha\! & \varphi_2^\alpha\!(\!\br_{\!2}\!)\alpha\! & \cdots & \varphi_1^\beta\!(\!\br_{\!2}\!)\beta\! & \varphi_2^\beta\!(\!\br_{\!2}\!)\beta\!   & \cdots \\ 
    \vdots                        & \vdots                        & \vdots & \vdots                      & \vdots                        & \vdots  \\
    \varphi_1^\alpha\!(\!\br_{\!N}\!)\alpha\! & \varphi_2^\alpha\!(\!\br_{\!N}\!)\alpha\! & \cdots & \varphi_1^\beta\!(\!\br_{\!N}\!)\beta\! & \varphi_2^\beta\!(\!\br_{\!N}\!)\beta\!   & \cdots
\end{array}\!\right|
\end{eqnarray}
The KS determinant yields the same electron density of the interacting system, with the following simple form ($N_\sigma$ being the number of $\sigma$-spin electrons),
\begin{equation}
\label{eq:rho_ks}
\rho_\sigma(\br)=\sum_{k=1}^{N_\sigma} |\varphi_k^\sigma(\br)|^2.
\end{equation}
The kinetic energy of the KS system can be calculated via
\begin{equation}
\label{eq:Ts_ks}
T_{\rm{s}}^{\rm{KS}}[\rho_\sigma]=-\frac{1}{2}\sum_{\sigma}^{\alpha,\beta}\sum_{k=1}^{N_\sigma} \langle \varphi_{k}^{\sigma} | \nabla^2 | \varphi_{k}^{\sigma} \rangle,
\end{equation}
which takes into account most of $T[\rho_\sigma]$ by resorting to the KS orbitals. With Eq. \ref{eq:Ts_ks}, the total energy can be expressed as
\begin{equation}
\label{eq:KS}
E_{\rm{tot}}[\rho_\sigma]=T_{\rm{s}}^{\rm{KS}}[\rho_\sigma]+E_{\rm{ext}}[\rho_\sigma]+E_{\rm{H}}[\rho_\sigma]+E_{\rm{XC}}^{\rm{KS}}[\rho_\sigma],
\end{equation}
and the unknown XC energy that needs to be obtained by approximation is much smaller,
\begin{equation}
\label{eq:XC_KS}
E_{\rm{XC}}^{\rm{KS}}[\rho_\sigma]=T[\rho_\sigma]-T_{\rm{s}}^{\rm{KS}}[\rho_\sigma]+E_{\rm{ee}}[\rho_\sigma]-E_{\rm{H}}[\rho_\sigma],
\end{equation}
which includes contributions from both the kinetic energy (i.e. $T[\rho_\sigma]-T_{\rm{s}}^{\rm{KS}}[\rho_\sigma]$) and the electron-electron interacting energy (i.e. $E_{\rm{ee}}[\rho_\sigma]-E_{\rm{H}}[\rho_\sigma]$).

As $T_{\rm{s}}^{\rm{KS}}[\rho_\sigma]$ is calculated by making use of the KS orbitals instead of the electron density, the ground-state energy in KS-DFT should be obtained by the minimization of $E_{\rm{tot}}[\rho_\sigma]$ with respect to $\{\varphi_p^\sigma\}$. In practice, the minimization can be achieved by solving the following KS equations \cite{KS1965}
\begin{equation}
\label{eq:eqKS}
\hat{h}^\sigma \varphi_p^\sigma(\br)=\varepsilon_p^\sigma \varphi_p^\sigma(\br),
\end{equation}
where the KS operator is defined by
\begin{equation}
\label{eq:opKS}
\hat{h}^\sigma = -\frac{1}{2}\nabla^2+v_{\rm{H}}(\br)+v_{\rm{XC}}^{{\rm{KS}},\sigma}(\br)+v_{\rm{ext}}(\br),
\end{equation}
and $v_{\rm{H}}(\br)$ and $v_{\rm{XC}}^{{\rm{KS}},\sigma}(\br)$ are the Coulomb and XC potentials, which are obtained via the functional derivatives $\frac{\delta E_{\rm{H}}[\rho_\sigma]}{\delta \rho_\sigma (\br)}$ and $\frac{\delta E_{\rm{XC}}^{\rm{KS}}[\rho_\sigma]}{\delta \rho_\sigma(\br)} $, respectively. The KS equations should be solved through the self-consistent field (SCF) calculation, as the KS operator depends on $\{\varphi_p^\sigma\}$ through the electron density of Eq. \ref{eq:rho_ks}.

In fact, the introduction of the KS orbitals not only improve the treatment of the kinetic energy with Eq. \ref{eq:Ts_ks}, but provides more electronic structure information in constructing functionals for better considering the XC effect. For example, glocal hybrid functionals \cite{Becke1993,Stephens1994,Ernzerhof1999,Adamo1999,
Xu2004,Xu2004jcp,Zhao2008}, local hybrid functionals \cite{Jaramillojcp2003,Arbuznikov2007cpl}, and different kinds of long-range corrected hybrid functionals \cite{Savin1996,Gill1996,Iikura2001,Yanai2004,LYPr,
Vydrov2006jcp,Chai2008,Heyd2004ajcp,Heyd038207,
Baer2010,Stein2010,Stein2009jcp,Peverati2011jpcl,Anderson2017jctc} that depend on the KS orbitals can achieve better accuracy, as compared to commonly used semi-local density functionals, for the prediction of various molecular properties. 


Despite these, KS-DFT still suffers from the static/strong correlation issue \cite{Burke2012jcp,Cohen2012cr,Becke2014jcp,Su2017arpc,Perdewe2021pnas}, manifested in the failure to predict the strong correlation energy. Furthermore, the KS determinant can produce incorrect densities for some systems, such as singlet oxygen \cite{Lee2019prl} and carbon atoms \cite{su2021unity}, and the SCF calculation may suffer from the convergence issue at the occurrence of (near-)degenerate states \cite{Rabuck1999jcp}. These problems occur when the single determinant description becomes inappropriate, so there is little sign that they can be completely solved by developing better approximate functionals. To this end, many efforts have been made to improve the description of strong correlation with multi-configuration DFT \cite{Leininger1997cpl,Li2014jctc,Gao2016jpcl,Chen2017jpcl}.

\subsection{Theory and functionals of HCKS} 
The HCKS theory provides a different idea to address the strong correlation issue, by generalizing the KS orbitals to hypercomplex number systems \cite{su2021unity}. The resulting high-dimension orbitals, termed the HCKS orbitals, take the following form
\begin{equation}
\label{eq:orb_hc}
\varphi_p^\sigma(\br)=\phi_{p}^{\sigma,0}(\br)+\sum_{\mu=1}^{n}\phi_{p}^{\sigma,\mu}(\br) e_\mu,
\end{equation}
where $\{e_1, e_2, \cdots, e_n\}$ are a basis of dimension $n$ in a Clifford algebra, such that \cite{HC1989,CA2019}
\begin{equation}
\label{eq:CA}
e_\mu^2=-1; e_\mu e_\nu=-e_\nu e_\mu,
\end{equation}
and $\{\phi_{p}^{\sigma,\mu}\}$ are a set of real functions. The conjugate hypercomplex of Eq \ref{eq:orb_hc} takes the form of
\begin{equation}
\label{eq:orb2_hc}
\bar{\varphi}_p^\sigma(\br)=\phi_{p}^{\sigma,0}(\br)-\sum_{\mu=1}^{n}\phi_{p}^{\sigma,\mu}(\br) e_\mu.
\end{equation}
The determinant constructed by the HCKS orbitals is called the HCKS determinant. 

Although the HCKS orbitals and HCKS determinant are hypercomplex, the corresponding density and kinetic energy are real, which can be obtained by inserting Eq \ref{eq:orb_hc} into Eqs \ref{eq:rho_ks} and \ref{eq:Ts_ks}, respectively. They are
\begin{equation}
\label{eq:rho2_hc}
\rho_\sigma(\br)=\sum_{k=1}^{N_\sigma}\sum_{\mu=0}^{n}[\phi_{k}^{\sigma,\mu}(\br)]^2,
\end{equation}
and
\begin{equation}
\label{eq:Ts2_hc}
T_{\rm{s}}^{\rm{HC}}[\rho_\sigma]=-\frac{1}{2}\sum_{\sigma}^{\alpha,\beta}\sum_{k=1}^{N_\sigma}\sum_{\mu=0}^{n} \langle\phi_{k}^{\sigma,\mu} |\nabla^2 | \phi_{k}^{\sigma,\mu} \rangle.
\end{equation}
The minimization of $E_{\rm{tot}}[\rho_\sigma]$ with respect to the HCKS orbitals leads to a set of high-dimension KS equations, named the HCKS equations,
\begin{eqnarray}
\label{eq:eq_hc}
\left\{ \begin{array}{c} 
{\hat{h}}^\sigma \phi_{p}^{\sigma,0}(\br)=\varepsilon_p^\sigma \phi_{p}^{\sigma,0}(\br)   \\
\hat{h}^\sigma \phi_{p}^{\sigma,1}(\br)=\varepsilon_p^\sigma \phi_{p}^{\sigma,1}(\br)  \\
\cdots   \\
{\hat{h}}^\sigma \phi_{p}^{\sigma,n}(\br)=\varepsilon_p^\sigma \phi_{p}^{\sigma,n}(\br). 
\end{array}\right.
\end{eqnarray}
It can be deduced from Eq \ref{eq:eq_hc} that all nonzero $\phi_{p}^{\sigma,\mu}$ for $\mu$ = 0 -- n are the eigenvectors of $\hat{h}^\sigma$ with the same eigenvalue $\varepsilon_p^\sigma$. When the eigenvector associated with $\varepsilon_p^\sigma$ is not degenerate, the components $\phi_{p}^{\sigma,\mu}$ in the $p$-th HCKS orbital are either 0 or different only by a constant factor; while for the degenerate case, degenerate eigenvectors and their mix can appear in different components of the $p$-th HCKS orbital, which can lead to fractional occupations for each degenerate eigenvectors.

Instead of solving Eq \ref{eq:eq_hc}, $\{\phi_{p}^{\sigma,\mu}\}$ can be further expanded on a set of orthonormal functions $\{\xi_p\}$ 
\begin{equation}
\label{eq:chi2phi}
\phi_{p}^{\sigma,\mu}(\br)=\sum_{q} \xi_q(\br)V_{pq}^{\sigma,\mu},
\end{equation}
where ${\rm{V}}^{\sigma,\mu}$ is a matrix associated with the $\mu$-th component of the HCKS orbitals. With this, the density of Eq \ref{eq:rho2_hc} becomes,
\begin{equation}
\label{eq:Drho}
\rho_\sigma(\br)=\sum_{p,q} \xi_p(\br) D_{pq}^{\sigma} \xi_q(\br),
\end{equation}
and ${\rm{D}}^\sigma$ is the density matrix on $\{\xi_p\}$ 
\begin{equation}
\label{eq:D_hc}
{\rm{D}}^\sigma=\sum_{\mu=0}^{n} {\rm{V}}^{\sigma,\mu T}{\rm{I}}^{N_\sigma}{\rm{V}}^{\sigma,\mu},
\end{equation}
where ${\rm{I}}^{N_\sigma}$ is a diagonal matrix, with the first $N_\sigma$ diagonal elements being 1 and the rest being 0. Because ${\rm{D}}^\sigma$ is symmetric, it can be diagonalized by an orthogonal matrix ${\rm{U}}^\sigma$,
\begin{equation}
\label{eq:ULU}
{\rm{D}}^\sigma={\rm{U}}^\sigma {\rm{\Lambda}}^\sigma {\rm{U}}^{\sigma T},
\end{equation}
where ${\rm{\Lambda}}^\sigma$ is a diagonal matrix, diag$(\lambda_1^\sigma, \lambda_2^\sigma, \cdots)$, and $\{\lambda_p^\sigma\}$ are the eigenvalues of ${\rm{D}}^\sigma$.
By inserting Eq \ref{eq:ULU}, the density of Eq \ref{eq:Drho} can be further formulated as
\begin{equation}
\label{eq:Vrho}
\rho_{\sigma}(\br) = \sum_{p=1}{\lambda}_{p}^{\sigma} |\chi_p^{\sigma}(\br)|^2.
\end{equation}
Here $\{\chi_k^\sigma\}$ are obtained via the unitary transformation on $\{\xi_k\}$: $\chi_p^{\sigma}(\br)=\sum_{q} \xi_q(\br)U_{qp}^{\sigma}$, which guarantees that $\{\chi_k^\sigma\}$ are orthonormal. The kinetic energy in HCKS is obtained by inserting Eqs \ref{eq:chi2phi}, \ref{eq:D_hc} and \ref{eq:ULU} into Eq \ref{eq:Ts2_hc}, which takes the form
\begin{equation}
\label{eq:VT}
T_{\rm{s}}^{\rm{HC}}[\rho_\sigma]=-\frac{1}{2}\sum_{\sigma}^{\alpha,\beta}\sum_{p} \lambda_{p}^{\sigma} \langle \chi_p^\sigma | \nabla^2 | \chi_p^{\sigma} \rangle.
\end{equation}
Eq. \ref{eq:VT} thus leads to the repartition of the total energy,
\begin{equation}
\label{eq:HCKS}
E_{\rm{tot}}[\rho_\sigma]=T_{\rm{s}}^{\rm{HC}}[\rho_\sigma]+E_{\rm{ext}}[\rho_\sigma]+E_{\rm{H}}[\rho_\sigma]+E_{\rm{XC}}^{\rm{HC}}[\rho_\sigma].
\end{equation}
As $T_{\rm{s}}^{\rm{HC}}[\rho_\sigma]$ and $E_{\rm{H}}[\rho_\sigma]$ take into account most of $T[\rho_\sigma]$ and $E_{\rm{ee}}[\rho_\sigma]$, respectively, the unknown XC energy in HCKS is small,
\begin{equation}
\label{eq:XC_HCKS}
E_{\rm{XC}}^{\rm{HC}}[\rho_\sigma]=T[\rho_\sigma]-T_{\rm{s}}^{\rm{HC}}[\rho_\sigma]+E_{\rm{ee}}[\rho_\sigma]-E_{\rm{H}}[\rho_\sigma],
\end{equation}
where the contribution from the kinetic energy (i.e. $T[\rho_\sigma]-T_{\rm{s}}[\rho_\sigma]$) is different from that of KS-DFT, while the contribution from the electron-electron interacting energy remains the same.

Thereby HCKS provides another way to handle the kinetic energy, by making use of the auxiliary orbitals $\{\chi_p^{\sigma}\}$ and their occupations $\{\lambda_p^\sigma\}$, accordingly, the ground-state energy has to be obtained by the minimization of $E_{\rm{tot}}[\rho_\sigma]$ with respect to both $\{\chi_p^\sigma\}$ and $\{\lambda_p^\sigma\}$, subject to some auxiliary constraints \cite{su2021unity}. The constrains on $\{\chi_p^\sigma\}$ is
\begin{equation}
\label{eq:cons_hco}
\langle \chi_p^\sigma | \chi_q^\sigma \rangle=\delta_{pq},
\end{equation}
while the constrains on $\{\lambda_p^\sigma\}$ is up to the dimension of the HCKS orbitals (i.e. $n+1$). In fact, the dimension of the HCKS orbitals determines the degree of restriction on $\{\lambda_p^\sigma\}$, which achieves correspondingly different HCKS methods with different definitions of $T_{\rm{s}}^{\rm{HC}}$ and $E_{\rm{XC}}^{\rm{HC}}$. When $n$ takes the minimum value (that is, $n=0$), $\{\lambda_p^\sigma\}$ are all integer (either 1 or 0), and HCKS reduces to KS. The constraints on $\{\lambda_p^\sigma\}$ need to be further explored for $n>0$, except when $n$ is infinite or $n+1\geq K$ for finite basis simulation ($K$ being the dimension of the basis set), the only constraints on $\{\lambda_p^\sigma\}$ are
\begin{equation}
\label{eq:cons_hcoo}
0\leq \lambda_p^\sigma \leq 1, \sum_{p} \lambda_p^\sigma=N_\sigma.
\end{equation}

Similar to the introduction of the KS orbitals in KS-DFT, the introduction of the auxiliary orbitals $\{\chi_p^\sigma\}$ and their occupations $\{\lambda_p^\sigma\}$ not only yields the electron density with Eq. \ref{eq:Vrho} and considers most of the kinetic energy with Eq. \ref{eq:VT}, but provides more flexibility in constructing functionals for better considering the XC effect. Unlike the KS orbitals whose occupations are only integer, $\{\lambda_p^\sigma\}$ can be either integer or fractional. This is interesting because $\{\lambda_p^\sigma\}$ are all (nearly) integer when the dynamic correlation effect is dominant; otherwise, some of $\{\lambda_p^\sigma\}$ could be fractional as a reflection of the static correlation. Thereby, $\{\chi_p^\sigma\}$ are the orbitals of hierarchical correlation (here termed HCOs), which contains distinct electronic information for the development of the XC functional toward better treatment of different correlations in HCKS.

The Hartree-Fock (HF) like exchange represents the simplest functional that depends on HCOs and their occupations, which takes the form of
\begin{equation}
\label{eq:HF_hc}
E_{\rm{X}}^{\rm{HF}}\!=\!-\!\frac{1}{2}\!\sum_\sigma^{\alpha,\beta}\!\sum_{p,q}\!\lambda_p^\sigma \lambda_q^\sigma\!\int\!\frac{\chi_p^\sigma(\br_1)\chi_p^{\sigma*}(\br_2)\chi_q^\sigma(\br_2)\chi_q^{\sigma*}(\br_1)}{r_{12}}d\br_1 d\br_2.
\end{equation}
Similarly, the long-range and short-range HF-like exchange functionals can be defined by including the error function ${\rm{erf}}(\omega r_{12})$ and the complementary error function ${\rm{erfc}}(\omega r_{12})$ in Eq \ref{eq:HF_hc}. With these, different global, local, and long-range corrected hybrid functionals can be developed and applied in HCKS as well.
In the following section, the semi-local functionals and the global hybrid functionals that combine semi-local functional with the HF-like exchange will be explored to gain insights into the functional development in HCKS.

\section{Results and discussion}

Here KS-DFT and HCKS in use of the same functionals are evaluated on the recent developed TS12 benchmark set \cite{Lee2019jcp}. This dataset includes 12 data points for triplet-singlet gaps of several atoms and diatomic molecules. The ground states of these systems are triplets, while the lowest singlet states are of biradical character. More tested details about TS12 can be found in refs \cite{Lee2019jcp,Lee2019prl}. Three nonempirical semi-local functionals are utilized, that is, PW92 \cite{PW91}, PBE \cite{PBE96}, and TPSS \cite{TPSS03}, which are functionals of local density approximations (LDA), generalized gradient approximations (GGA), and meta-GGA (mGGA), respectively. In addition, the combination of the three semi-local functionals with different amounts of HF-like exchange is also examined to achieve further insights into the HCKS methods. All calculations were performed using a local modified version of the NWChem package \cite{NWChem}, where the occupation optimization is handled in the explicit-by-implicit (EBI) manner \cite{Yao2021jpcl}.
The basis set of aug-cc-pVQZ \cite{Dunning1989jcp,Kendall1992jcp} is used for all the calculations.

\begin{table*}
\caption{Triplet-singlet gaps $\Delta E_{{\rm {T-S}}}$ ($=E_{{\rm {S}}}-E_{{\rm {T}}}$) (kcal/mol) of various atoms and diatoms from the TS12 benchmark set
\cite{Lee2019jcp}. Both spin-restricted and -unrestricted KS-DFT (denoted as RKS and UKS, respectively) and HCKS are tested in use of LDA (PW92), GGA (PBE), and mGGA (TPSS), where the GGA results are obtained from Ref \cite{su2021unity}.  Spin-unrestricted calculations are performed for triplet states, while spin-restricted calculations are performed for singlet states except UKS. 
Experimental data are used as reference. MSE and MAE stand for mean sign error
and mean absolute error, respectively. }
\label{tab:tpss} %
\begin{tabular}{ccccccccccccc}
\hline 
 &  & \multicolumn{3}{c}{RKS} &  & \multicolumn{3}{c}{UKS} &  & \multicolumn{3}{c}{HCKS}\tabularnewline
\cline{3-5} \cline{4-5} \cline{5-5} \cline{7-9} \cline{8-9} \cline{9-9} \cline{11-13} \cline{12-13} \cline{13-13} 
 & Expt.  & LDA & GGA & mGGA  &  & LDA & GGA & mGGA  &  & LDA & GGA & mGGA \tabularnewline
\hline 
C  & 29.14  & 42.89 & 43.31 & 42.28  &  & 14.98 & 8.60 & 5.22  &  & 27.71 & 31.72 & 40.59 \tabularnewline
NF  & 34.32  & 43.58 & 45.36 & 46.09  &  & 14.89 & 10.75 & 9.52  &  & 29.91 & 34.15 & 40.61 \tabularnewline
NH  & 35.93  & 51.13 & 53.36 & 56.03  &  & 17.59 & 12.75 & 14.11  &  & 35.68 & 41.20 & 50.22 \tabularnewline
NO$^{-}$  & 17.30  & 25.93 & 26.97 & 27.79  &  & 9.00 & 6.66 & 6.12  &  & 17.87 & 20.46 & 24.50 \tabularnewline
O$_{2}$  & 22.64  & 36.87 & 37.56 & 37.51  &  & 11.84 & 8.66 & 7.38  &  & 23.80 & 26.50 & 31.03 \tabularnewline
O  & 45.37  & 64.54 & 66.56 & 68.80  &  & 21.53 & 16.20 & 17.28  &  & 36.39 & 43.80 & 55.63 \tabularnewline
PF  & 20.27  & 28.44 & 32.00 & 32.67  &  & 8.27 & 6.99 & 5.98  &  & 18.06 & 22.83 & 26.11 \tabularnewline
PH  & 21.90  & 30.73 & 34.60 & 35.65  &  & 9.10 & 7.76 & 7.57  &  & 19.73 & 25.08 & 28.95 \tabularnewline
S$_{2}$  & 13.44  & 20.35 & 22.44 & 23.08  &  & 5.49 & 4.65 & 4.16  &  & 11.98 & 14.79 & 17.21 \tabularnewline
S  & 26.41  & 36.96 & 40.95 & 42.06  &  & 10.46 & 9.03 & 8.85  &  & 17.83 & 23.87 & 29.29 \tabularnewline
Si  & 18.01  & 25.55 & 28.74 & 28.64  &  & 7.81 & 5.94 & 4.39  &  & 14.74 & 19.75 & 23.69 \tabularnewline
SO  & 18.16  & 26.14 & 28.08 & 28.52  &  & 7.56 & 5.99 & 5.16  &  & 16.08 & 19.13 & 22.24 \tabularnewline
\hline 
MSE  &  & 10.85 & 13.09 & 13.85  &  & -13.70 & -16.57 & -17.26  &  & -2.76 & 1.70 & 7.27 \tabularnewline
MAE  &  & 10.85 & 13.09 & 13.85  &  & 13.70 & 16.57 & 17.26  &  & 3.05 & 2.41 & 7.27 \tabularnewline
\hline 
\end{tabular}
\end{table*}

The test results of the semi-local functionals can be found in Tab \ref{tab:tpss}. As can be seen, the performances of KS-DFT and HCKS are quite different in use of the same functionals. The spin-restricted KS-DFT (denoted as RKS) systematically overestimates the triplet-singlet gaps of the 12 systems in TS12. Due to the theoretical limitation, RKS can only provide a closed-shell solution for any singlet state, with all orbitals either doubly occupied or empty. Thereby, it cannot capture the multi-reference nature of the singlet biradicals, resulting in overestimation of the energies of the lowest singlet states. Note that even though RKS can correctly predict the zero spin-densities ($\rho_\alpha-\rho_\beta$) for these singlets, it breaks the space symmetry of the total densities. Unlike RKS, the spin-unrestricted KS-DFT (denoted as UKS) can predict the biradical character by breaking the spin symmetry, with two unpaired electrons of opposite spins confined in different physical region. However, the destruction of spin symmetry leads to the over-relaxation of the orbitals occupied by the unpaired opposite electrons, thus the energies of the singlet states as well as the triplet-singlet gaps are systematically underestimated. Similar results and discussions about KS-DFT can be found in ref. \cite{Lee2019prl,su2021unity}. In contrast, HCKS allows dynamically varying fractional occupations to capture the multi-reference nature of strong correlation. Thereby, HCKS significantly improves the predicted triplet-singlet gaps, with much smaller errors as compared to KS-DFT. Nevertheless, the approximate functional for the XC energy still has a considerable influence on the performance of HCKS.  For example, TPSS systematically overestimates the triplet-singlet gaps, with both MSE and MAE being 7.27 kcal/mol. The MAE can be reduced to 3.05 and 2.41 kcal/mol for PW92 and PBE, respectively. Therefore, functionals developed specifically for HCKS are essential to further improve the performance of HCKS. 

\begin{figure}[htbp]
 \centering
 \includegraphics[width=1\linewidth]{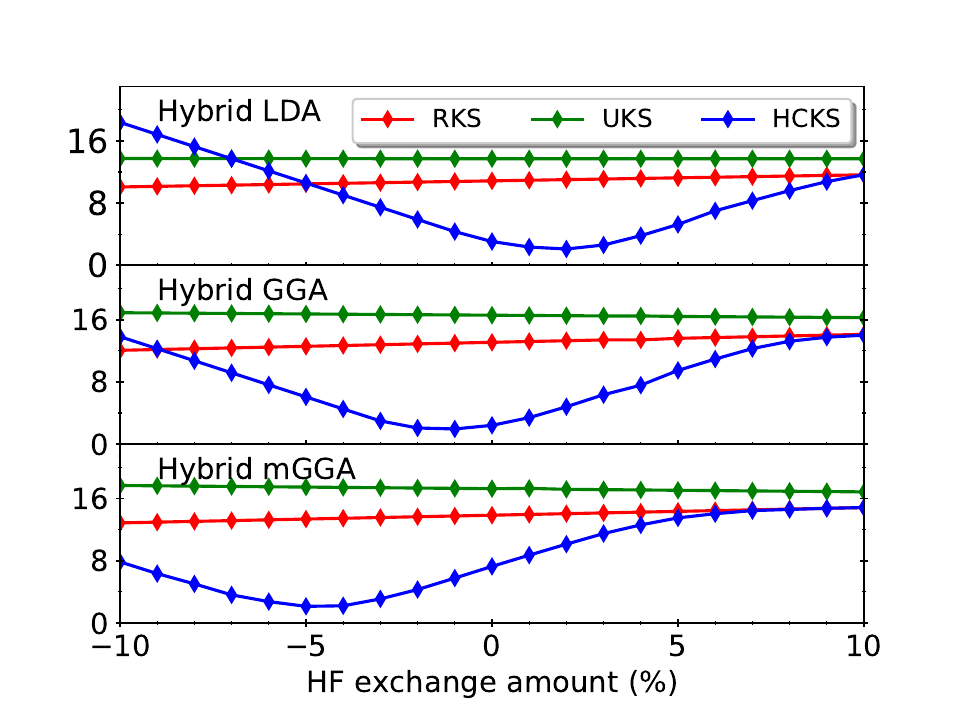}
 \caption{Mean absolute errors (MAEs) of triplet-singlet gaps $\Delta E_{\rm{T-S}}$ ($=E_{\rm{S}}-E_{\rm{T}}$) by KS-DFT and HCKS in use of hybrid functionals with different amount of HF exchange ($c_{\rm{X}}^{\rm{HF}}$). Both spin-restricted and -unrestricted KS-DFT, denoted as RKS and UKS, are evaluated. 
 The tested functionals take the from: $E_{\rm{XC}}=c_{\rm{X}}^{\rm{HF}}E_{\rm{X}}^{\rm{HF}}+(1-c_{\rm{X}}^{\rm{HF}})E_{\rm{X}}^{\rm{DFA}}+E_{\rm{C}}^{\rm{DFA}}$, which combines LDA (PW92), GGA (PBE), and mGGA (TPSS) with the HF-like exchange. Spin-unrestricted calculations are performed for triplet states, while spin-restricted calculations are performed for singlet states except UKS. 
 The test set is the TS12 benchmark set \cite{Lee2019jcp}. All energies are in kcal/mol.}
 \label{fig:hybrids}
\end{figure}

In order to gain more understanding of the functional development toward better description of strong correlation in HCKS, the combination of different semi-local functionals with the HF-like exchange that depends on HCOs and their occupations are explored here. To this end, the triplet-singlet gaps of TS12 are calculated by KS-DFT and HCKS in use of the hybrid functionals with varying amount of HF exchange ($c_{\rm{X}}^{\rm{HF}}$), i.e. $E_{\rm{XC}}=c_{\rm{X}}^{\rm{HF}}E_{\rm{X}}^{\rm{HF}}+(1-c_{\rm{X}}^{\rm{HF}})E_{\rm{X}}^{\rm{DFA}}+E_{\rm{C}}^{\rm{DFA}}$. It can be clearly seen from Fig \ref{fig:hybrids} that the inclusion of the HF-like exchange considerably affects the performance of HCKS, while LDA, GGA, mGGA require different amounts of the HF-like exchange to reach the minimal errors for this test set. 
It has been proved that the LDA exchange functional includes not only the exchange energy but also the static correlation effect \cite{NICHOLAS2001}; GGA and mGGA improve the treatment of exchange energy, including less static correlation energy \cite{ACDH}. Therefore, the hybrid GGA and mGGA functionals in HCKS require a large amount of semi-local exchange functionals to bring in sufficient static correlation for properly describing singlet biradicals. In particular, the systematic error of mGGA@HCKS can be considerably reduced if a negative fraction of HF-like exchange is included, and the best performance can be obtained when $c_{\rm{X}}^{\rm{HF}}=-5\%$, with the MSE and MAE reduced from 7.27 and 7.27 kcal/mol to -0.30 and 2.13 kcal/mol, respectively. Note that the functionals tested here are used to gain some understanding of the approximate functionals of HCKS, more tests are needed before they are used in other calculations. 

For better understanding, the hybrid functionals can be rewritten in the from:  $E_{\rm{XC}}=E_{\rm{XC}}^{\rm{DFA}}+\Delta E_{\rm{XC}}$, where $\Delta E_{\rm{XC}}=c_{\rm{X}}^{\rm{HF}}(E_{\rm{X}}^{\rm{HF}}-E_{\rm{X}}^{\rm{DFA}})$ is considered as a correction to the LDA, GGA, and mGGA functionals, and the XC functionals used in KS-DFT and HCKS can be continuously changed by adjusting $c_{\rm{X}}^{\rm{HF}}$ to include different amounts of $\Delta E_{\rm{XC}}$. By observing Fig \ref{fig:hybrids}, it can be found that the correction $\Delta E_{\rm{XC}}$ has little impact on KS-DFT. The systematical errors of RKS and UKS rarely change for different $c_{\rm{X}}^{\rm{HF}}$, and the same results can be obtained by the hybrid functionals based on LDA, GGA, and mGGA. This thus indicates that the description of strong correlation poses a great challenge to DFT within the frameworks of KS-DFT, while HCKS creates new possibilities for the development and evaluation of approximate functionals toward better consideration of strong correlation. 

\section{Conclusion and outlook}

This work discussed approximate functionals in HCKS, seeking to gain in-depth insights into the development of approximate functionals through the comparison between KS-DFT and HCKS. Unlike the KS orbitals that are integer occupied, HCKS allows dynamically varying fractional occupations to capture the multi-reference nature of strong correlation. These orbitals of hierarchical correlation (HCOs) thus contain distinct electronic information for the development of the XC functional toward better treatment of different correlations in HCKS. The test on the triplet-singlet gaps of the TS12 benchmark set shows that the performance of HCKS depends on the choice of approximate functionals. In particular, the systematic errors of semi-local functionals can be effectively reduced by including appropriate amount of the HCO-dependent HF exchange. Unlike HCKS, KS-DFT shows large systematic errors on the triplet-singlet gaps, worse still, the errors rarely change when the amount of HF-like exchange is continuously adjusted in functionals. Therefore, HCKS has the great potential to address the strong correlation issue of KS-DFT, and further development of functionals that depend on HCOs and their occupations is essential for the application of HCKS in strongly correlated systems. The insights and advances gained in this work will be helpful in the development and evaluation of approximation functionals in DFT.

Support from the National Natural Science Foundation of China (Grants No. 22073049 and No. 22122303), the Natural Science Foundation of Tianjin City (20JCQNJC01760), and Fundamental Research Funds for the Central Universities (Nankai University: No. 63206008) is appreciated.

\bibliographystyle{aip}
\bibliography{ref}

\end{document}